\documentclass[10pt,letterpaper]{article} 
\usepackage{opex3}
\usepackage{graphicx}
\usepackage{amsmath}
\usepackage{amssymb}
\usepackage{mathrsfs}
\usepackage{amsfonts}
\usepackage{dsfont}
\usepackage{dcolumn}% Align table columns on decimal point
\usepackage{bm}% bold math
\usepackage{booktabs}
\usepackage{textcomp}
\usepackage{cite}

\everymath{\displaystyle}
\newcommand{\di}{\mathrm{d}}

\newcommand{\br}{\bm{r}}

\newcommand{\bn}{\bm{n}}

\newcommand{\bk}{\bm{k}}

\newcommand{\bp}{\bm{p}}
\newcommand{\bj}{\bm{j}}

\newcommand{\bE}{\bm{E}}
\newcommand{\bB}{\bm{B}}

\newcommand{\bA}{\bm{A}}

\newcommand{\bnabla}{\bm{\nabla}}
\newcommand{\beps}{\bm{\epsilon}}
\newcommand{\bbeta}{\bm{\beta}}

\newcommand{\bEt}{\widetilde{\bm{E}}}
\newcommand{\bBt}{\widetilde{\bm{B}}}
\newcommand{\bAt}{\widetilde{\bm{A}}}

\newcommand{\Et}{\widetilde{E}}
\newcommand{\At}{\widetilde{A}}

\newcommand{\thetat}{\widetilde{\theta}}
\newcommand{\alphat}{\widetilde{\alpha}}

\providecommand{\abs}[1]{\lvert#1\rvert}

\newcommand{\brp}{\bm{r}_\perp}
\newcommand{\bkp}{\bm{k}_\perp}
\begin{document}

%%%new commands
\newcommand{\barr}{\begin{eqnarray}}
\newcommand{\earr}{\end{eqnarray}}
%column vector
\newcommand{\vett}[2]{\left[
\begin{array}{c}
#1\\#2
\end{array}
\right]}
%line vector
\newcommand{\Vett}[2]{\left[
\begin{array}{cc}
#1 &#2
\end{array}
\right]}
%matrix
\newcommand{\mat}[4]{\left[
\begin{array}{cc}
#1 & #2 \\ #3 & #4 \\
\end{array}
\right]}
%beam
\newcommand{\beam}[1]{\mathbf{#1}(\bm{r})}

\title{Surface angular momentum of light beams}

\author{Marco Ornigotti$^{1}$ and Andrea Aiello$^{2,3,*}$}
\address{$^1$Institute of Applied Physics, Friedrich-Schiller University, Jena, Max-Wien Platz 1, 07743 Jena, Germany\\
$^2$ Max Planck Institute for the Science of Light, G$\ddot{u}$nther-Scharowsky-Strasse 1/Bau24, 91058 Erlangen,
Germany\\ 
$^3$Institute for Optics, Information and Photonics, University of Erlangen-Nuernberg, Staudtstrasse 7/B2, 91058 Erlangen, Germany}
\email{* andrea.aiello@mpl.mpg.de}
\date{\today}
\begin{abstract}

Traditionally, the angular momentum of light is calculated for ``bullet-like'' electromagnetic wave packets, although in actual optical experiments ``pencil-like'' beams of light are more commonly used. The fact that a wave packet is bounded  transversely and longitudinally while a beam has, in principle, an infinite extent along the direction of propagation,  renders incomplete the textbook calculation of the spin/orbital separation of the  angular momentum of a light beam. 
In this work we demonstrate that a novel, extra surface part must be added in order to preserve the gauge invariance of the optical angular momentum per unit length.
The impact of this extra term is quantified by means of two examples: a  Laguerre-Gaussian  and a Bessel beam, both circularly polarized.  
\end{abstract}

\ocis{(260.0260) Physical Optics; (260.6042) Singular Optics; (260.2110) Electromagnetic Optics.}

\section{Introduction}
In his \emph{Treatise of Electricity and Magnetism} \cite{maxwell}, Maxwell pointed out that the electromagnetic field carries both energy and momentum, and that the momentum can have both linear and angular contributions. Today it is common knowledge that the  angular momentum (AM) of light can be thought of having two contributions: a spin part associated to polarization (firstly theoretically investigated by Poynting \cite{poynting} and then experimentally demonstrated by Beth \cite{beth}), and an orbital part associated with the spatial distribution of the field, as first recognised by Darwin \cite{darwin}. About sixty years later, thanks to the seminal work of Allen and Woerdman \cite{allen1}, the topic of  AM of light experienced a new renaissance, producing a considerable amount of literature on the subject, including theoretical discussions about the separation of AM in  spin and orbital parts for optical beams \cite{barnett2,barnett,Enk1,Enk2,Berry,Li,Bliokh2010,Barnett3,Birula}, the spin-Hall effect of light \cite{SHEL3, OLandrea, SHEL1,SHEL2}, the geometric spin-Hall effect of light \cite{beck, GSHEL}, and a huge variety of experimental applications such as optical tweezers, spatial light modulators and vortex beams. A satisfactory review of the development of the field of optical  AM of light can be found in Refs. \cite{review,phase_front,libro_andrea}.

The AM of light has been frequently separated, by many authors, into a spin part and an orbital part (see, e.g., \cite{haus} for a didactic presentation of the subject). As emphasized by Birula\&Birula \cite{Birula} such separation can be either gauge dependent or gauge invariant. The latter result is achieved, according to Darwin \cite{darwin},  by expressing the optical AM in terms of the Fourier transform of the electromagnetic fields. On the contrary, the gauge dependent separation is obtained, traditionally, when in the expression of the optical AM
\begin{align}\label{JAM}
\pmb{\mathcal{J}} =\epsilon_0 \int \bm{r} \times \left( \bm{E} \times \bm{B} \right) \di^3 r ,
\end{align}
%
%where $\di^3 r = \di x \, \di y \, \di z$ denotes volume integration, 
the magnetic field $\bB$ is replaced by the curl of the vector potential $\bA$, namely $\bB = \bm{\nabla} \times \bA$. After partial integration  Eq. \eqref{JAM} takes the split form \cite{CTan,Joerg}
\begin{align}\label{JAM2}
\pmb{\mathcal{J}} = \epsilon_0 \int  \bm{E} \times \bm{A}  \, \di^3 r + \epsilon_0 \int \sum_{\xi } E_\xi 
\left( \bm{r} \times \bnabla \right) A_\xi \, \di^3 r - \epsilon_0 \int \sum_{\xi } \nabla_\xi \left(
E_\xi \, \bm{r} \times \bA \right)  \, \di^3 r,
\end{align}
where $\xi \in \{ x,y,z\}$ and the first and the second integral represent the spin part and the orbital part, respectively, of the optical AM. The third integrals is usually omitted in the textbook expressions of $\pmb{\mathcal{J}}$ because, as consequence of the Gauss' theorem, it will be identically zero if the fields vanish sufficiently quickly as $\abs{\br} \to \infty$. This requirement, as pointed out by Crichton\&Marston \cite{3div} and Nieminen \textit{et al.} \cite{Nieminen}, is often understood while its fulfillment should be carefully checked case by case.   For example, the electric and magnetic fields of an optical ``bullet-like'' wave packet with a finite  transverse and longitudinal (with respect to the direction of propagation) extent, by definition vanish as $\abs{\br} \to \infty$ and the third term in Eq. \eqref{JAM2} can be safely neglected. However, this is no longer true for a ``pencil-like'' beam of light whose span  along the direction of propagation is virtually infinite. This simple but important fact was already noticed by Barnett\&Allen \cite{barnett2} (see Eqs. (3.15-18) in their paper) who, nevertheless, 
simply classified the third integral in Eq. \eqref{JAM2} as a nonparaxial contribution to the AM \emph{per unit length} of the beam and did not investigate its physical content.

In this paper we thoroughly investigate the physical nature of the latter term in Eq. \eqref{JAM2} that we call the ``Su\emph{rface} A\emph{ngular} M\emph{omentum}'' (SuAM) of light. At the fundamental level, we demonstrate that even for an  optical beam with a small angular spread around the direction of propagation (paraxial beams), the SuAM term must be retained in order to guarantee the gauge invariance of the theory. From an experimental point of view, we show that SuAM
arises whenever the intensity of an optical beam  is recorded by a planar detector which, in practice, performs a two dimensional integration over a cross section of the beam perpendicular to its direction of propagation. As this is the case occurring in the majority of experimental optical setups, we believe that the impact of our new results may be significant. All our conclusions are valid for light beams propagating in free space.

The structure of this Article is as follows. In Sec. II we first calculate  the total  AM density $\bj(\bm{r})$ of a monochromatic beam of light described in terms of the vector potential $\bm{A}$.
Then, by following the standard procedure, we decompose $\bj(\bm{r})$ into the sum of three terms: the well-known orbital and spin  AM parts and the third term, leading to the SuAM. Next, we integrate  $\bj(\bm{r})$ upon a two-dimensional surface at $z=const.$, mimicking the detection process, thus obtaining the total  AM per unit length $\bm{J}$. This unravels the nature of the  SuAM term and how it manifests  in the actual detection of the  AM of a light beam. For the sake of completeness, we conclude this section by calculating  the explicit form of the  SuAM for the cases of Laguerre-Gaussian and Bessel beams.
In Sec. III we calculate again $\bm{J}$ but, this time,  in a gauge invariant manner by using the angular spectrum representation of the electric and magnetic fields. Then, by comparing the expressions for $\bm{J}$ obtained in Secs. II and III we show that the SuAM term cannot be neglected even for paraxial optical beams.  
Finally, we draw our conclusions.
\section{Theory of  SuAM}
To begin with, consider a monochromatic beam of light  with angular frequency $\omega$ whose  real electric and magnetic fields $\pmb{\mathcal{E}}(\bm{r},t)$ and $\pmb{\mathcal{B}}(\bm{r},t)$ are expressed in terms of the complex field amplitudes $\bm{E}(\bm{r})$ and $\bm{B}(\bm{r})$ as $\pmb{\mathcal{E}}(\bm{r},t)=\text{Re}[\bm{E}(\bm{r})\exp(-i \omega t )]$ and $\pmb{\mathcal{B}}(\bm{r},t)=\text{Re}[\bm{B}(\bm{r})\exp(-i \omega t )]$.
Following van Enk and Nienhuis \cite{Enk2} 
we write the cycle-averaged linear momentum, angular momentum and energy densities, respectively,
\begin{subequations}\label{pANDjDENSITIES}
\begin{align}
\bp(\br) = & \; \frac{\varepsilon_0}{2}\,\text{Re}\left( \bE^* \times \bB \right), \\
\bj(\br) = & \; \frac{\varepsilon_0}{2}\,\br\times\text{Re}\left( \bE^* \times \bB  \right), \\  
u(\br) = & \; \frac{\varepsilon_0}{4}\, \left( \bE \cdot \bE^* + c^2 \bB \cdot \bB^* \right),  
\end{align}
\end{subequations}
 where  $\bm{E}$ and    $\bm{B}$ are written in terms of the transverse vector potential $\bm{A}(\bm{r})$ as \cite{CTan}
\begin{align}\label{EandB}
\bm{E} = i \omega \bm{A}\quad \text{and} \quad \bm{B} = \bm{\nabla} \times \bm{A}\quad \text{with} \quad \bm{\nabla} \cdot \bm{A} =0.
\end{align}
By using Eqs. (\ref{pANDjDENSITIES}-\ref{EandB})
 we can express the  AM  densities $\bm{j}(\bm{r})$ as \cite{3div,Humblet}:
%
%\begin{subequations}\label{eq40}
%
\begin{align}\label{eq40}
\bj(\br) = & \; \frac{\omega  \, \epsilon_0}{2} \,\mathrm{Re}\bigg\{\bA^*\cdot\big(- i\br\times\bm{\nabla}\big)\bA - i \bA^*\times\bA  +i\sum_{\xi\in \{x,y,z\}} \frac{\partial}{\partial \xi}\left[A_\xi^*(\br \times \bA)\right]\bigg\}.
\end{align}
%
%\end{subequations}
%
 According to \cite{3div}, the first term in Eq. \eqref{eq40} represents the orbital part of the  AM and the second term gives the spin  AM of light. The third term is a three-divergence that vanishes when integrated over a volume $V$ whose surface boundary $S$ is far enough from the field sources to give $\left. \beam{A} \right|_S\approx \bm{0}$. However, as  noticed in   \cite{3div}, the latter relation should be checked whenever the decomposition  \eqref{eq40} is used. In the remainder of this Article, we will show some consequences of the failure of the relation  $\left. \beam{A} \right|_S\approx \bm{0}$.

Haus and Pan \cite{haus}  pointed out that for a monochromatic beam of light of angular frequency $\omega$ propagating in the $z$ direction, the classical optics analogous of the helicity $\pm \hbar$ of a photon, namely the projection of the spin  AM along the direction of the linear momentum, is given by  the ratio $\hat{\bm{e}}_z\cdot \bm{J}/U=\pm 1/\omega $,
where  $\bm{J}$ and  $U$  denote the cycle-averaged  AM and energy \emph{per unit length}:
\begin{align}\label{pANDjANDu}
%
%\;
\bm{J} =  \int  \bm{j}(\bm{r}) \, \text{d}^2 r, \qquad U =  \int u(\bm{r}) \, \text{d}^2 r,
\end{align}
where $\text{d}^2 r = \text{d} x \, \text{d} y$ and the integrations extend over all the plane $xy$ at $z=const.$ 
It should be noticed that for a monochromatic optical beam that can be represented by a superposition of homogeneous plane waves  solely (namely, without contributions from evanescent waves) both $\bm{J}$ and $U$ do not depend upon the longitudinal coordinate $z$ and, therefore, are conserved during free propagation.
Henceforth,  we shall use over line symbol to denote normalization according to the rule
\begin{align}\label{sigmaVett}
\overline{\bm{J}} =\frac{\bm{J}}{U/\omega}.
\end{align}
This definition of helicity is  consistent with the physical picture of a beam of light as observed on a laboratory bench, where it travels across plates, lenses and other optical devices before eventually shining the surface of a \emph{planar} detector. Last but not least, it should be noticed that the calculations by Haus and Pan \cite{haus} are valid not only in the \emph{strict} paraxial limit where the fields are purely transverse with respect to the propagation axis $z$, but also when first-order corrections (in the sense of Lax \emph{et al.}, \cite{lax}), leading to small longitudinal fields components, are accounted for.  

As a consequence of the surface integration, a beam-like field ``feels'' the contribution given by the $z$-part of the latter term in Eq. \eqref{eq40}, namely $\partial_z[A_z^*(\bm{r}\times\bm{A})]$, because such a term is \emph{not} integrated when one calculates $\bm{J}$. In practice, since $\int\partial_\xi[A_\xi^*(\bm{r}\times\bm{A})]\, \text{d}^2 r=\partial_z\int A_z^*(\bm{r}\times\bm{A})\, \text{d}^2 r$,
we can write the  SuAM contribution in the following form:
\begin{align}\label{SuAM}
\bm{J}^{\text{surf}}=\frac{\partial}{\partial z}\int \mathrm{Re}\Big\{iA_z^*(\bm{r}\times\bm{A})\Big\}\, \text{d}^2 r,
\end{align}
where $\mathrm{Re}\left\{iA_z^*(\bm{r}\times\bm{A})\right\}=i\left[z(\bm{A}^*\times\bm{A})-\bm{r}\cdot(\bm{A}^*\times\bm{A})\hat{\bm{e}}_z\right]/2$.
From Eq. \eqref{SuAM} it appears that the $\bm{J}^{\text{surf}}$ depends both on the potential vector $\bm{A}$ and the position vector $\bm{r}$, thus revealing a seemingly hybrid spin/orbital nature. 
Now, we can calculate $\bm{J}^{\text{surf}}$ from Eq. \eqref{SuAM}  by writing
 the vector potential as a superposition of \emph{homogeneous} plane waves
\begin{align}
\bA(\br)= & \; \frac{1}{2 \pi} \int \bAt(\bk_\perp) \exp\left(i \br_\perp \cdot \bk_\perp + i z \,k_z \right) \di^2 k , \label{AS}
\end{align}
where $\bAt(\bk_\perp): \, \bk \cdot \bAt(\bk_\perp)  =0$ is the so-called angular spectrum of the field evaluated at $z=0$ \cite{mandelWolf} and  $\bk = (k_x, k_y,k_z )$ is the wave vector with $k = \abs{\bk}$ and $\omega =  k \,c$. For the sake of consistency, 
henceforth we shall use the following notation: $\brp =( x,y ,0)$, $\bkp = (k_x, k_y,0 )$,  $\br_\perp \cdot \bk_\perp = x \,k_x + y \, k_y$, $k_\perp = \abs{\bkp}$ and $\di^2 k  = \di k_x \, \di k_y$. 
By definition, $k_z $ is not an independent variable and can be fixed to the non-negative value  $k_z = +(k^2 - k_\perp^2)^{1/2}\geq 0$, which is appropriate for the forward homogeneous fields considered here. 
For a circularly polarized optical beam, the angular spectrum can be chosen of the form
\begin{align}\label{eq400}
\bAt(\bkp) = & \; \At (\bkp) \beps_\perp(\sigma,\bkp), \nonumber \\
= & \; \abs{\At (\bkp)} \exp \big[i \, \alphat(\bkp) \big] \beps_\perp(\sigma,\bkp),
\end{align}
where the real-valued amplitude $\abs{\At (\bkp)}$ and phase $ \alphat(\bkp)$ determines the spatial profile of the beam and  $\beps_\perp(\sigma,\bkp)$ is the complex-valued polarization vector defined as:
\begin{align}\label{eq410}
\beps_\perp(\sigma,\bkp) = & \; \bn_\sigma - \bk \left(\bk \cdot \bn_\sigma \right )/k^2   \nonumber \\
= & \;  - \bk \times \left(\bk \times \bn_\sigma \right)/k^2 .
\end{align}
In the equation above $\bn_\sigma = (\hat{\bm{e}}_x+ i \sigma \hat{\bm{e}}_y)/\sqrt{2}$ with $\sigma = \pm 1$, represents  the wave vector-\emph{independent} configuration-space helicity assigned in the (global) laboratory reference frame $\{\hat{\bm{e}}_x,\hat{\bm{e}}_y,\hat{\bm{e}}_z\}$.
By definition $\beps_\perp(\sigma,\bkp)$ is not normalized; however, since its norm
\begin{align}
\beps_\perp^*(\sigma,\bkp) \cdot \beps_\perp(\sigma,\bkp) = & \;1 - \abs{\bk \cdot \bn_\sigma}^2/k^2   \nonumber \\
= & \;  1- \frac{k_\perp^2}{2 k^2}   \nonumber \\
\equiv & \; 1  - \vartheta^2 ,
\end{align}
is polarization-independent, we can assume that the normalization factor is tacitly contained in the expression of $\At (\bkp)$. Here,  the dimensionless parameter $0 \leq \vartheta<1/\sqrt{2}$:
\begin{align}
\vartheta^2 = \frac{k_x^2 + k_y^2}{2 k^2},
\end{align}
defines the angular spread of the beam around the direction of propagation $z$, and  the  paraxial regime is characterized by angular spectra such that the condition $\vartheta \ll 1$ is satisfied.
At this point, by substituting Eq. \eqref{AS} into Eq. \eqref{eq40} and using Eqs. (\ref{eq400}-\ref{eq410}) one obtains after a straightforward calculation the orbital part
\begin{subequations}\label{orbA}
\begin{align}
J_x^\text{orb} = & \; -\frac{\epsilon_0 \omega}{2} \int \abs{\bAt}^2 \left[ \sigma \,\frac{k_x k_z}{k^2+k_z^2} +  k_z\frac{\partial \alphat}{\partial k_y} \right] \di^2 k , \\
J_y^\text{orb} = & \; -\frac{\epsilon_0 \omega}{2} \int \abs{\bAt}^2 \left[ \sigma \,\frac{k_y k_z}{k^2+k_z^2} -  k_z\frac{\partial \alphat}{\partial k_y} \right] \di^2 k , \\
J_z^\text{orb} = & \;  \frac{\epsilon_0 \omega}{2} \int \abs{\bAt}^2  \, \left[ \sigma \,  \frac{\vartheta^2}{1 - \vartheta^2}  +  \left(k_x \frac{\partial \alphat}{\partial k_y} -k_y \frac{\partial \alphat}{\partial k_x} \right)\right] \di^2 k ,
\end{align}
\end{subequations}
the spin part
\begin{subequations}\label{spinA}
\begin{align}
J_x^\text{spin} = & \; \frac{\epsilon_0 \omega}{2} \, 2 \sigma \, \int \abs{\bAt}^2  \, \frac{k_x k_z}{k^2 + k_z^2} \, \di^2 k , \\
J_y^\text{spin} = & \; \frac{\epsilon_0 \omega}{2} \, 2 \sigma \, \int \abs{\bAt}^2  \, \frac{k_y k_z}{k^2+ k_z^2} \, \di^2 k , \\
J_z^\text{spin} = & \; \frac{\epsilon_0 \omega}{2} \,  \sigma \, \int \abs{\bAt}^2 \, \frac{1 - 2 \vartheta^2}{1- \vartheta^2} \, \di^2 k , 
\end{align}
\end{subequations}
 and the surface part of the optical AM per unit of length:
\begin{subequations}\label{surfA}
\begin{align}
J_x^\text{surf} = & \; -\frac{\epsilon_0 \omega}{2} \, \sigma \, \int \abs{\bAt}^2  \, \frac{k_x k_z}{k^2 + k_z^2}  \, \di^2 k , \label{surfA1}\\
J_y^\text{surf} = & \; -\frac{\epsilon_0 \omega}{2} \, \sigma \, \int \abs{\bAt}^2  \, \frac{k_y k_z}{k^2 + k_z^2}  \, \di^2 k , \label{surfA2} \\
J_z^\text{surf} = & \; \frac{\epsilon_0 \omega}{2} \, \sigma \, \int \abs{\bAt}^2  \, \frac{\vartheta^2}{1- \vartheta^2}  \,  \di^2 k . \label{surfA3}
\end{align}
\end{subequations}
In addition, the energy per unit length is given by
\begin{align}\label{energyA}
\frac{U}{\omega} = \frac{\epsilon_0 \, \omega}{2}  \int \abs{\bAt}^2  \, \di^2 k .
\end{align}
Note that all the quantities (\ref{orbA}-\ref{energyA}) are $z$-independent, so they \emph{cannot} vanish if one would perform an additional integration along the axis $z$ to obtain the AMs in a given, finite, volume.

The explicit form for $\bm{J}^{\text{surf}}$ is the first main result of this Article. What does it represent?
First of all, it should be noticed that at this stage the angular spectrum $\bAt (\bkp)$ is perfectly general and may represent, for example, an electromagnetic beam with arbitrary  orbital AM.
 The  $z$-component of the integrand in Eq. \eqref{surfA3} is proportional to $\vartheta^2/(1 - \vartheta^2)$ as noticed already in \cite{barnett2}. This contribution is $O(\vartheta^2)$ and therefore disappears in the strict paraxial limit where such terms are not retained \cite{haus}. In order to understand better how $\bm{J}^{\text{surf}}$ affects the total  AM per unit length we sum the three contribution and obtain 
\begin{align}
J_z^\text{orb} +J_z^\text{spin}+J_z^\text{surf}=   \frac{\epsilon_0 \omega}{2} \int \abs{\bAt}^2  \, \left[  \,  \frac{\sigma}{1 - \vartheta^2}  +  \left(k_x \frac{\partial \alphat}{\partial k_y} -k_y \frac{\partial \alphat}{\partial k_x} \right)\right] \di^2 k. \label{JtotA}
\end{align}
It is instructive to compare Eq. \eqref{JtotA} with Eq. (38) in  \cite{barnett} where Barnett calculates the ratio between the cycle-averaged  angular momentum density flux $\mathcal{M}_{zz}$ and the energy density  flux $\mathcal{F}$ through the $xy$ plane,  namely $\mathcal{M}_{zz}/\mathcal{F}$.
In our case, for a circularly polarized beam with helicity $\sigma$ and by using Eq. \eqref{AS}, it follows that 
\begin{align}\label{normF}
\mathcal{M}_{zz}=   \frac{\epsilon_0 \, c^2}{2} \int \abs{\bAt}^2 \, k_z \,   \left[  \,  \sigma  +  \left(k_x \frac{\partial \alphat}{\partial k_y} -k_y \frac{\partial \alphat}{\partial k_x} \right)\right] \di^2 k,
\end{align}
and
\begin{align}\label{norm}
\mathcal{F}=   \frac{c^2 \epsilon_0 \, \omega}{2} \int \abs{\bAt}^2 \, k_z \, \di^2 k.
\end{align}
The main difference between Eq. \eqref{JtotA} and Eq. \eqref{normF} resides, apart from the trivial factor $c^2/\omega$, in the extra multiplicative term $k_z = k (1- 2 \vartheta^2)^{1/2} \simeq k (1-  \vartheta^2)$ in the integrand. Of course, in the paraxial limit $k_z \to k$ and the two expressions coincides. There is, however, a profound conceptual and practical difference between the energy density $U$ and the energy density flux $\mathcal{F}$ (Poynting vector flux). The physical  quantity that is actually measured by common detection devices as CCD detectors, photographic plates, photoresists, etc.,
is the time-averaged value of the \emph{scalar} energy density (namely, loosely speaking, the number of photons in the unit volume) integrated over the detector surface,  rather than the Poynting \emph{vector} flux  
\cite{Loudon,Braat,Braat2}. This distinction between scalar and vector quantities becomes crucial when measuring spin-dependent non-paraxial optical phenomena as, e.g., the geometric spin Hall effect of light \cite{GSHEL}. In fact, the standard theory of photo-detection (see, e.g., sec. 4.11 of \cite{Loudon} and chap. III of \cite{CTan}) shows that the \emph{observable intensity} of the electromagnetic field is proportional to the probability of observing a photoionization in a phototube detector which is given by the expectation value of the electric-field energy density operator $\hat{\bm{E}}^{(-)}(\bm{r},t) \cdot \hat{\bm{E}}^{(+)}(\bm{r},t)$. As shown in   \cite{Loudon}, only when a light beam is at least approximately paraxial the observable intensity 
coincides with the flux across the detector surface of the Poynting vector operator  $\hat{\bm{I}}(\bm{r},t)= \epsilon_0 c^2 \left\{ \hat{\bm{E}}^{(-)}(\bm{r},t) \times \hat{\bm{B}}^{(+)}(\bm{r},t)- \hat{\bm{B}}^{(-)}(\bm{r},t) \times \hat{\bm{E}}^{(+)}(\bm{r},t) \right\}$. Therefore, in general, is the electromagnetic energy density and not the Poynting vector flux that matters in ordinary photodetection.

It should be noticed that both $\bm{J}^{\text{surf}}$ and $\bm{J}^{\text{spin}}$ are proportional to the helicity $\sigma$ of the incident beam. However, this should not be interpreted as a signature of the spin nature of the SuAM, since such a dependence also appears in spin-to-orbit conversion phenomena \cite{Grier}. 
From Eqs. (\ref{spinA}-\ref{surfA}) it follows that
\begin{align}\label{indep}
J_z^{\text{spin}} + J_z^ {\text{surf}}=  \frac{\epsilon_0 \omega}{2} \, \sigma \, \int \abs{\bAt}^2  \,  \di^2 k , 
\end{align}
from which, by using  Eqs. \eqref{sigmaVett} and  \eqref{energyA},  we obtain
\begin{align}\label{indep2}
\overline{J_z^{\text{spin}} } +\overline{J_z^{\text{surf}}} = \sigma.
\end{align}
Equation \eqref{indep2} shows that adding the $z$-component of the  SuAM term to the $z$-component of the spin AM per unit length ensures 
 that the total ``spin'' AM per unit length along the propagation axis $z$ is \emph{exactly} equal to $\sigma$ times the energy per unit length of the beam, the latter being completely arbitrary. This our second main result.

\subsection{Examples}In the remainder of this section, we calculate explicitly the SuAM for  Laguerre-Gaussian and Bessel beams to investigate how orbital AM  affects $\overline{\bm{J}^{\text{surf}}}$.
 Consider first the case of a Laguerre-Gaussian  beam of waist $w_0$, angular aperture  $\theta_0=2/(k w_0)$ and with azimuthal and radial mode indexes $\ell$ and $p$, respectively \cite{Siegman}. In this case, from Eqs. (\ref{spinA}-\ref{surfA})  we obtain
\begin{align}\label{gb SuAM}
\overline{\bm{J}^{\text{surf}}} =  \sigma \frac{\beta}{1-\beta}\hat{\bm{e}}_z, \quad \overline{\bm{J}^{\text{spin}}} =  \sigma \frac{1-2\beta}{1-\beta}\hat{\bm{e}}_z
,
\end{align}
where ${\beta = (\theta_0/2)^2(2p+|\ell| + 1 )}$ \cite{NoteBarnett}.
 Although $\beta$ depends on $|\ell|$ such a dependence is not a  consequence of the orbital AM of the beam (this would lead to a dependence from the signed quantity $\ell$), but it comes from the radial dependence of the beam profile. For a fundamental Gaussian beam $\ell=0=p $ and $\beta \to \theta_0^2/4$. In this case, for $\theta_0 \ll 1$, the small angle approximation furnishes $\overline{J_z^{\text{spin}}}/  \sigma \simeq 1-\theta_0^2/4 $, in agreement with \cite{Nieminen}.

Next, we consider the case of a $\ell$th-order Bessel beam that, in the cylindrical polar coordinates $(r,\phi,z)$, takes the form $E(r,\phi,z)=J_\ell(r k \sin \vartheta_0)\exp{(i\ell\phi)}\exp{(iz k \cos \vartheta_0)}$,
where  $\vartheta_0$ is the angular aperture of the characteristic Bessel cone. A straightforward calculation shows that Eqs. \eqref{gb SuAM} are still valid providing that $\beta = \sin^2 \vartheta_0 /2$.
Note that for both Laguerre-Gaussian and Bessel \emph{paraxial} beams with $\theta_0=\vartheta_0 \ll 1$, one has $\overline{\bm{J}^{\text{surf}}} \sim O(\theta_0^2)$, namely $\overline{\bm{J}^{\text{surf}}}$ becomes negligible.
\begin{figure}[!tbp]
\begin{center}
\includegraphics[width=10truecm]{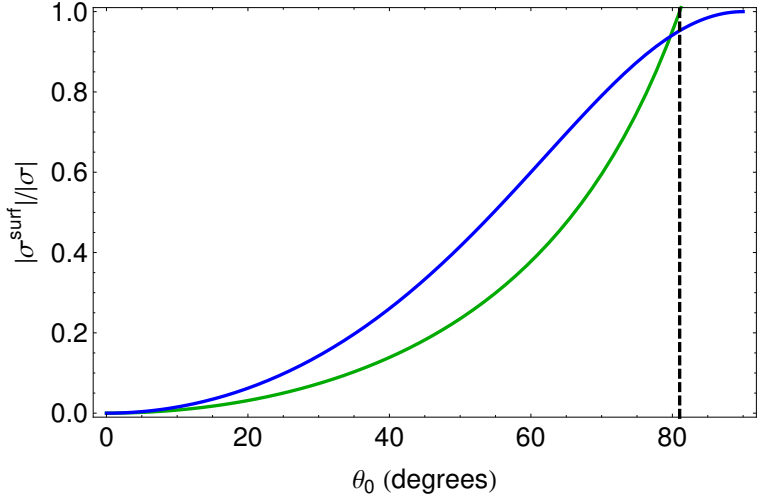}
\caption{ SuAM for a circularly polarized fundamental Gaussian beam (green line) and Bessel beam (blue line). 
At  $\theta_0 = \theta_c \equiv \sqrt{2} \; \text{rad} \, \simeq 81^\circ$ the tails of the Gaussian angular spectrum distribution $|\bAt(\bm{k}_\perp)|^2 \propto \exp[- 2 (k_x^2 + k_y^2)/\theta_0^2]$  becomes no longer negligible  for $k_x^2 + k_y^2 \geq k^2$ where  evanescent waves occur \cite{mandelWolf}. The angular spectrum representation Eq. \eqref{AS} with $k_z =+(k^2 - k_x^2 - k_y^2)^{1/2}\in \mathbb{R}$ does not account for evanescent waves, therefore it  breaks down for $\theta_0> \theta_c$. This critical value is marked by a  dashed vertical line. Such a problem does not occur for a Bessel beam because in this case the angular spectrum does not possess tails but is sharply peaked about $\vartheta_0$. 
}
\label{fig1}
\end{center}
\end{figure}
The absolute value  of the ``surface helicity'' $\sigma^\text{surf} \equiv \overline{\bm{J}^{\text{surf}}}\cdot\hat{\bm{e}}_z$,  is depicted in Fig. \ref{fig1} for both the case of a fundamental Gaussian beam (green solid line) and of a Bessel beam (blue solid line) for $\vartheta_0=\theta_0\in[0,\pi/2]$.
In the paraxial domain $\sigma^{\text{surf}}$ approaches zero and its contribution can be neglected for $\theta_0\ll 1$. 
However, for non paraxial values of $\theta_0$, $|\sigma^{\text{surf}}|$  becomes comparable with the standard helicity value  $|\sigma|=1$ \cite{haus}.
\section{Gauge invariant calculation}
In this section we calculate again the AM per unit length for a monochromatic beam of light. However, this time we will not use the vector potential and, thus, our calculation will be manifestly gauge invariant. Having this goal in mind, we write the electromagnetic fields in their (homogeneous) angular spectrum form as
\begin{subequations}\label{FieldsEB}
\begin{align}
\bE(\br)= & \; \frac{1}{2 \pi} \int \bEt(\bk_\perp)\exp\left(i \br_\perp \cdot \bk_\perp + i z \,k_z \right) \di^2 k , \label{FieldsE}\\
\bB(\br)= & \; \frac{1}{2 \pi} \int \bBt(\bk_\perp)\exp\left(i \br_\perp \cdot \bk_\perp + i z \, k_z \right) \di^2 k ,\label{FieldsB}
\end{align}
\end{subequations}
where $\bEt$ and $\bBt$ fulfill the following relations stemming from Maxwell equations:
\begin{subequations}\label{MaxCond}
\begin{align}
\bk \cdot \bEt  = & \; 0,  \\
\bk \cdot  \bBt  = & \; 0,   \\
 c \, \bBt  = & \; \frac{\bk}{k} \times  \bEt.  \label{MaxCond3}
\end{align}
\end{subequations}
As before, for a circularly polarized beam of light we can choose
\begin{align}
\bEt(\bkp)  =  & \; \Et (\bkp) \beps_\perp(\sigma,\bkp) \nonumber \\
 =  & \;\abs{\Et (\bkp)} \exp \big[i \, \thetat(\bkp) \big] \beps_\perp(\sigma,\bkp)
\end{align}
and from Eq. \eqref{MaxCond3} it follows that $c \, \bBt(\bkp)  =  \Et (\bkp) \, \bbeta_\perp(\sigma,\bkp)$, where 
\begin{align}\label{polB}
\bbeta_\perp(\sigma,\bkp)  = & \; \bk \times \beps_\perp(\sigma,\bkp)/k \nonumber \\
= & \; \left(\bk \times \bn_\sigma \right)/k .  
\end{align}
Once again, we assume that the normalization factor for the polarization vector $\beps_\perp(\sigma,\bkp)$ is contained in the complex-valued amplitude $\Et (\bkp)$.

By substituting Eqs. (\ref{FieldsEB}-\ref{polB}) into Eqs. \eqref{pANDjDENSITIES} we obtain, after a straightforward calculation, the following expressions for the energy per unit length
\begin{align}
\frac{U}{\omega} = \frac{\epsilon_0}{2 \omega}  \int \abs{\bEt}^2    \, \di^2 k ,
\end{align}
for the Poynting vector
\begin{align}
\bm{S} = & \;  c^2 \frac{\epsilon_0}{2 \omega} \int \abs{\bEt}^2     \,  \bm{k}  \, \di^2 k ,
\end{align}
and for the AM per unit length:
\begin{subequations}\label{Jgi}
\begin{align}
J_x = & \; \frac{-\epsilon_0}{2 \omega} \int \abs{\bEt}^2    \, k_z \frac{\partial \, \thetat}{\partial k_y} \, \di^2 k ,\label{Jgix} \\
J_y = & \; \frac{\epsilon_0}{2 \omega}  \int \abs{\bEt}^2    \, k_z  \frac{\partial \, \thetat}{\partial k_x} \, \di^2 k , \label{Jgiy} \\
J_z = & \;  \frac{\epsilon_0}{2 \omega} \int \abs{\bEt}^2     \, \left[  \,  \frac{\sigma}{1 - \vartheta^2  } + 
 \left(k_x \frac{\partial \, \thetat}{\partial k_y} -k_y \frac{\partial \, \thetat}{\partial k_x} \right) \right] \di^2 k  . \label{Jgiz}
\end{align}
\end{subequations}
By comparing Eq. \eqref{Jgiz} with the expression of the AM flux below 
\begin{align}
\mathcal{M}_{zz} = & \; c^2 \, \frac{\epsilon_0}{2 \omega^2} \int \abs{\bEt}^2     \, k_z   \left[ \sigma + \left(k_x \frac{\partial \, \thetat}{\partial k_y} -k_y \frac{\partial \, \thetat}{\partial k_x} \right)\right] \di^2 k ,
\end{align}
one may be tempted  to identify the term proportional to $\sigma$ in Eq.  \eqref{Jgiz} with the spin part of the AM per unit length. However, such a term differs from the expression of $J_z^\text{spin}$ given by Eq.  \eqref{spinA}. Moreover, from Eqs. (\ref{orbA}-\ref{spinA}) it follows that
\begin{align}\label{JorbPlusSpin}
J_z^{\text{orb}} + J_z^ {\text{spin}}=  \frac{\epsilon_0 \omega}{2} \int \abs{\bAt}^2  \, \left[ \sigma   +  \left(k_x \frac{\partial \alphat}{\partial k_y} -k_y \frac{\partial \alphat}{\partial k_x} \right)\right] \di^2 k ,
\end{align}
which does not coincides with  Eq. \eqref{Jgiz}. However, if we add $J_z^{\text{surf}}$ to Eq. \eqref{JorbPlusSpin} we obtain the result given by Eq.  \eqref{JtotA} that we rewrite below:
\begin{align}
J_z^\text{orb} +J_z^\text{spin}+J_z^\text{surf}=   \frac{\epsilon_0 \omega}{2} \int \abs{\bAt}^2  \, \left[  \,  \frac{\sigma}{1 - \vartheta^2}  +  \left(k_x \frac{\partial \alphat}{\partial k_y} -k_y \frac{\partial \alphat}{\partial k_x} \right)\right] \di^2 k. \label{JtotA2}
\end{align}
Finally, the two expressions given by Eq. \eqref{Jgiz} and Eq. \eqref{JtotA2} coincide (apart from the trivial factor $\omega$ linking the electric field and the vector potential amplitudes $\bEt$ and $\bAt$, respectively). An analogous result is found for the other two components  $J_x$ and $J_y$. It should be noted that Eq. \eqref{JtotA2} is independent from $z$, thus indicating that the total (spin plus orbital plus surface) AM per unit of length is conserved during the free propagation of the beam.

In summary, we can recollect the results obtained in this section as follows: By using the angular spectrum representation for the electromagnetic fields, we have calculated the AM per unit length $\bm{J}$ for an arbitrary beam of light, \emph{either paraxial or non paraxial}. Then, we compared this result  given by Eqs. \eqref{Jgi} with the expressions (\ref{orbA}-\ref{surfA}) of $\bm{J}^\text{orb}$, $\bm{J}^\text{spin}$ and $\bm{J}^\text{surf}$ and we found that
\begin{align}
\bm{J}= \bm{J}^\text{orb} + \bm{J}^\text{spin}+ \bm{J}^\text{surf}.
\end{align}
The equation above unambiguously shows that the gauge invariant expression of $\bm{J}$ on the left side can be obtained via the gauge dependent calculation on the right side, only when the SuAM is accounted for, irrespective of the paraxial or non paraxial nature of the beam of light.
\section{Conclusions}
One of the key aspects of this work, is the illustration of the interplay between paraxial approximation, AM of light  and gauge invariance. The actual situation can be schematically illustrated as follows. In the simplest electromagnetic field, the plane wave, the electric and magnetic fields are strictly perpendicular to the direction of propagation of the wave. Differently, for a beam of finite waist, a nonzero component of the electric field $E_z$ parallel to the direction of propagation $z$ of the beam, is unavoidable \cite{Erikson}. This component, although small with respect to $E_x$ and $E_y$ in collimated beams, is responsible for the difference between \textit{A}) linear and angular momentum fluxes from one side, and \textit{B}) energy density and AM per unit of length from the other side. In fact, as shown in \cite{barnett}, both fluxes depends only upon the transverse fields components $E_x, \, E_y$ and $B_x, \, B_y$. On the opposite, it can be shown that the energy density and the AM per unit of length, receive a contribution from $E_z$ and $B_z$, as well \cite{GSHEL}. What quantities are more physically meaningful, \textit{A}) or \textit{B})? There is not a definite answer to this question. In principle, the fluxes obey more general conservation laws and therefore, from a theoretical point of view, are more appealing. On the other hand, the quantities more commonly measured  in standard laboratories, are the energy density and the AM per unit of length. In practice, it will be the experimental configuration at hand to set the choice between \textit{A}) and \textit{B}), because
spin, orbital and total AM of light may be measured by means of  different techniques \cite{libro_andrea}. Many detection apparatuses yield the ratio between the angular momentum per unit length and the energy density per unit length. 
Typical examples thereof are Stokes parameter measurements in paraxial polarization optics (the third Stokes parameter, usually denotes $s_3$, gives a measure of $\sigma$; see, e.g., \cite{3div} and Sec. \textbf{6.2} of \cite{mandelWolf}) and interferometric methods in single photon AM detection \cite{Leach}. In all these cases, the energy density of light per unit length is measured by means of commonplace photo-detectors as CCD cameras, photographic plates, photoresists, etc. and the quantities calculated in this work are the relevant ones.

How gauge invariance enters this discussion? In the paraxial approximations  the quantities \textit{A}) and \textit{B}) coincide and  the spin/orbital separation of the AM is manifestly gauge-invariant in both cases. However, beyond paraxial approximation only the fluxes  \textit{A}) remain explicitly gauge-invariant, as shown by Barnett \cite{barnett}. Oppositely, gauge-invariance is no longer guaranteed for the AM per unit of length outside paraxial approximation.  In fact, 
in this paper we have demonstrated that the  textbook expression for the angular momentum per unit length of a beam of light, containing a spin and an orbital part solely, is not complete and that a third part must be included in order to preserve the gauge invariance of the theory. We call this new term the su\emph{rface} AM (SuAM). This quantity is derived considering the virtually infinite extent of the beam along the direction of propagation.
We believe that our results may have a relevant conceptual impact upon the very lively and timely research field about  AM of light \cite{prlBliokh,peter}.

\section*{Acknowledgment}
%
%
%
%\begin{verbatim} \ack = NJP command for acknowledgment \end{verbatim}
%
We are grateful to Konstantin Bliokh for fruitful and stimulating discussions. 


\begin{thebibliography}{99}
%
%
\bibitem{maxwell}J. C. Maxwell, \emph{A Treatise on Electricity and Magnetism}, (Dover, 1954).
%
\bibitem{poynting}J. H. Poynting, ``The Wave Motion of a Revolving Shaft, and a Suggestion as to the Angular Momentum in a Beam of Circularly Polarised Light'',  Proc. R. Soc. Lond. A Ser. A \textbf{82}, 560-567 (1909).
%
\bibitem{beth}R. A. Beth, ``Mechanical Detection and Measurement of the Angular Momentum of Light'',  Phys. Rev. \textbf{50}, 115-125 (1936).
%
\bibitem{darwin}C. G. Darwin, ``Notes on the Theory of Radiation'',  Proc. R. Soc. Lond. A \textbf{136}, 36-52 (1932).
%
\bibitem{allen1}L. Allen, M. W. Beijersbergen, R. J. C. Spreeuw, and J. P. Woerdman, ``Orbital angular momentum of light and the transformation of Laguerre-Gaussian laser modes'',  Phys. Rev. A \textbf{45}, 8185-8189 (1992).
%
\bibitem{barnett2}S. M. Barnett and L. Allen, ``Orbital angular momentum and nonparaxial light beams'',  Opt. Comm. 
\textbf{110}, 670-678 (1994).
%
\bibitem{Enk1}  S. J. van Enk and G. Nienhuis, ``Spin and Orbital Angular Momentum of Photons'',   Europhys. Lett. \textbf{25}, 497-501 (1994).
%
\bibitem{Enk2}  S. J. van Enk and G. Nienhuis, ``Commutation Rules and Eigenvalues of Spin and Orbital Angular Momentum of Radiation Fields'',  J. Mod. Opt. \textbf{41}, 963-977 (1994).
%
\bibitem{barnett}S. M. Barnett, ``Optical angular-momentum flux'',  J. Opt. B: Quantum Semiclass. Opt. \textbf{4}, S7-S16 (2002).
%
\bibitem{Berry} M. V. Berry, ``Optical currents'',  J. Opt. A: Pure Appl. Opt. \textbf{11}, 094001 (2009).
%
\bibitem{Li}  C.-F. Li, ``Spin and orbital angular momentum of a class of nonparaxial light beams having a globally defined polarization'',   Phys. Rev. A \textbf{80}, 063814 (2009).
%
\bibitem{Bliokh2010} K. Y. Bliokh, M. A. Alonso, E. A. Ostrovskaya, and A. Aiello, ``Angular momenta and spin-orbit interaction of nonparaxial light in free space'',  
Phys. Rev. A \textbf{82}, 063825 (2010).
%
%
\bibitem{Barnett3} S. M. Barnett, ``Rotation of electromagnetic fields and the nature of optical angular momentum'',  J. Mod. Opt. \textbf{57}, 1339-1343 (2010).
%
\bibitem{Birula} I. Bialynicki-Birula and Z.  Bialynicki-Birula, ``Canonical separation of angular momentum of light into its orbital and spin parts'',  J. Opt. \textbf{13}, 064014 (2011).
%
\bibitem{SHEL3}O. Hosten and P. Kwiat, ``Observation of the Spin Hall Effect of Light via Weak Measurements'',  \emph{Science} \textbf{319}, 787-790 (2008).
%
\bibitem{OLandrea}A. Aiello and J. P. Woerdman, ``Role of beam propagation in GoosÐHŠnchen and ImbertÐFedorov shifts'',  Opt. Lett. \textbf{33}, 1437-1439 (2008).
%
\bibitem{SHEL1}M. Onoda, S. Murakami, and N. Nagaosa, ``Hall Effect of Light'',  Phys. Rev. Lett. \textbf{93}, 083901 (2004).
%
\bibitem{SHEL2}K. Y. Bliokh and Y. P. Bliokh, ``Conservation of Angular Momentum, Transverse Shift, and Spin Hall Effect in Reflection and Refraction of an Electromagnetic Wave Packet'',  Phys. Rev. Lett. \textbf{96},  073903 (2006).
%
\bibitem{beck}A. Y. Bekshaev, ``Oblique section of a paraxial light beam: criteria for azimuthal energy flow and orbital angular momentum'',  J. Opt. A: Pure Appl. Opt. \textbf{11}, 094003 (2009).
%
\bibitem{GSHEL}A. Aiello, N. Lindlein, Ch. Marquardt, and G. Leuchs, ``Transverse Angular Momentum and Geometric Spin Hall Effect of Light'',  Phys. Rev. Lett. \textbf{103}, 100401 (2009).
%
\bibitem{review}S. Franke-Arnold, L. Allen, and M. Padgett, ``Advances in optical angular momentum'',   Laser \& Photon. Rev. \textbf{2}, 299-313 (2008).
%
\bibitem{phase_front}\emph{Optical Angular Momentum}, edited by L. Allen, S. M. Barnett, and M. J. Padgett (Institute of Physics, Bristol, England, 2003).
%
\bibitem{libro_andrea}K. Y. Bliokh, A.  Aiello, and M. 
Alonso, ``Spin-orbit interactions of light in isotropic media,'' in The Angular
Momentum of Light, D. L. Andrews and M. Babiker, eds. (Cambridge University, 2012), pp. 174-245.
%
\bibitem{haus}H. A. Haus and J. L. Pan, ``Photon spin and the paraxial wave equation'',   Am. J. Phys. \textbf{61}, 818-821 (1993).
%
\bibitem{CTan} C. Cohen-Tannoudji, J. Dupont-Roc, G. Grynberg, \emph{Photons \& Atoms}, (Wiley-VCH, Weinheim, 2004), Chap. I.
%
\bibitem{Joerg} J. B. G\"{o}tte and S. M. Barnett,
``Light beams carrying orbital angular momentum,'' in The Angular Momentum of
Light, D. L. Andrews and M. Babiker, eds. (Cambridge University, 2012), pp. 1-30.
%
\bibitem{3div}J. H. Crichton and P. L. Marston, ``The measurable distinction between the spin and orbital angular momenta of electromagnetic radiation'',   Electronic Journal of Differential Equations, http://www.emis.de/journals/EJDE/, pp.37-50 (2000).
%
\bibitem{Nieminen} T. A. Nieminen, A. B. Stilgoe, N. R. Heckenberg, and A. Rubinsztein-Dunlop, ``Angular momentum of a strongly focused Gaussian beam'',  J. Opt. A: Pure Appl. Opt. \textbf{10}, 115005 (2008).
%
\bibitem{Humblet} J. Humblet, ``Sur le moment d'impulsion d'une onde \'electromagn\`etique'',  Physica (Utrecht)  \textbf{10},  585-603 (1943).
%
\bibitem{lax} M. Lax, W. H. and Louisell, and W. B. McKnight, ``From Maxwell to paraxial wave optics'',  Phys. Rev. A \textbf{11}, 1365-1370 (1975).
%
\bibitem{mandelWolf}L. Mandel and E. Wolf, \emph{Optical Coherence and Quantum Optics}, (Cambridge University Press, 1995).
%
\bibitem{Loudon} R. Loudon, \emph{The Quantum Theory of Light} 3rd ed., (Oxford University Press, 2000).
%
\bibitem{Braat} J. J. M. Braat, P. Dirksen,  A. J. E. M. Janssen, and A. S. van de Nes, ``Extended Nijboer-Zernike representation of the vector field in the focal region of an aberrated high-aperture optical system'',  J. Opt. Soc. Am. A \textbf{20}, 2281 (2003).
%
\bibitem{Braat2} J. J. M. Braat, S. van Haver, A.J.E.M. Janssen, and P. Dirksen, ``Energy and momentum flux in a high-numerical-aperture beam using the extended Nijboer-Zernike diffraction formalism'',  J.  Eur. Opt. Soc. - Rapid  \textbf{2}, 07032 (2007).
%
\bibitem{Grier} D. B. Ruffner and D. G. Grier, ``Optical Forces and Torques in Nonuniform Beams of Light'',  Phys. Rev. Lett. \textbf{108}, 173602 (2012).
%
\bibitem{Siegman} A. E. Siegman, \emph{Lasers}, (University Science Books, 1986).
%
\bibitem{NoteBarnett}
The same ``correction'' factor $\beta$ was found in  \cite{barnett2}. However, the expression (4.4) in \cite{barnett2}, when evaluated for $l=0=p$  in the paraxial limit $1/(2 k z_R) = \theta_0^2 /4 \ll 1$ gives $\mathscr{J}_z/\mathscr{E} \simeq (\sigma_z/\omega)(1 + \theta_0^2/4)$ instead of $\mathscr{J}_z/\mathscr{E} \simeq (\sigma_z/\omega)(1 - \theta_0^2/4)$, as  given by Eq. (8) in \cite{Nieminen}. The difference between the two results  resides in the fact that different types of beams and geometries (planar versus spherical) are considered. 
%
\bibitem{Erikson} W. L. Erikson and S. Singh, ``Polarization properties of Maxwell-Gaussian laser beams'' Phys. Rev. A \textbf{49}, 5778-5786 (1994).
%
%
\bibitem{Leach} J. Leach, J. Courtial, K. Skeldon, S. M. Barnett, S. Franke-Arnold, and M. J. Padgett, ``Interferometric Methods to Measure Orbital and Spin, or the Total Angular Momentum of a Single Photon'', Phys. Rev. Lett. \textbf{92}, 013601 (2004).
%
%
\bibitem{prlBliokh} O. G. Rodr\'{\i}guez-Herrera, D. Lara, K. Y. Bliokh, E. A. Ostrovskaya, and C. Dainty, ``Optical Nanoprobing via Spin-Orbit Interaction of Light'',  Phys. Rev. Lett. \textbf{104}, 253601 (2010).
%
\bibitem{peter}P. Banzer, U. Peschel, S. Quabis, and G. Leuchs, ``On the experimental investigation of the electric and magnetic response of a single nano-structure'',  Opt. Expr. \textbf{18}, 10905-10923 (2010).
%
%
%
%
\end{thebibliography}
\end{document}